\documentclass[referee]{jfm}
\usepackage{natbib}
\usepackage{graphicx,epsfig}
\usepackage{upmath}
\usepackage{amssymb}
\usepackage{amsmath}
\usepackage{amstext}
\usepackage[hang]{subfigure}

\def\be{\begin{equation}}
\def\ee{\end{equation}}
\def\ben{\begin{equation*}}
\def\een{\end{equation*}}

\renewcommand{\thesubfigure}{\thefigure}    %{\thefigure.\arabic{subfigure}}
\makeatletter
   \renewcommand{\@thesubfigure}{\thesubfigure:\space}
   \renewcommand{\p@subfigure}{}
\makeatother

\title[Conditionally Cubic-Gaussian Stochastic Lagrangian Model]{A Conditionally Cubic-Gaussian Stochastic Lagrangian Model
for Acceleration in Isotropic Turbulence}
\author[A.~G.~Lamorgese$^1$, S.~B.~Pope$^1$, P.~K.~Yeung$^2$ and B.~L.~Sawford$^3$]
{A.\ns G.\ns L\ls A\ls M\ls O\ls R\ls G\ls E\ls S\ls E$^1$,\ns S.\ns B.\ns P\ls O\ls P\ls E$^1$,\\
P.\ns K.\ns Y\ls E\ls U\ls N\ls G$^2$ \and B.\ns L.\ns S\ls A\ls W\ls F\ls O\ls R\ls D$^3$}

\affiliation{$^1$Sibley School of Mechanical \& Aerospace Engineering, \vspace{-0.3cm} \\ 
 Cornell University, Ithaca, N.Y. 14853-7501, USA\\[\affilskip]
$^2$School of Aerospace Engineering, Georgia Institute of Technology, \vspace{-0.3cm} \\ 
 270 Ferst Drive, Atlanta, Georgia,  30332-0150, USA \\[\affilskip]
$^3$Dept.~of Mechanical Engineering, Monash University,   \vspace{-0.3cm}\\ 
Clayton Campus, Wellington Road, Clayton, VIC 3800, Australia}

\pubyear{2006}
\volume{}
\pagerange{}
\date{\today}
\begin{document}

\maketitle

\begin{abstract}

\noindent

The modelling of fluid particle accelerations in homogeneous, isotropic turbulence in terms 
of second-order stochastic models for the Lagrangian velocity is considered.
The basis for the Reynolds model (A.~M.~Reynolds, \textit{Phys.~Rev.~Lett.}~$\mathbf{91}(8)$, 084503 (2003))
is reviewed and examined by reference to DNS data. In particular, we show DNS data
that support stochastic modelling of the logarithm of pseudo-dissipation 
as an Ornstein-Uhlenbeck process (Pope and Chen 1990)
and reveal non-Gaussianity of the conditional acceleration PDF.
The DNS data are used to construct a simple stochastic model that is exactly consistent with Gaussian
velocity and conditionally cubic-Gaussian acceleration statistics.
This model captures the effects of intermittency of dissipation on acceleration and the
conditional dependence of acceleration on pseudo-dissipation (which differs from that predicted by the refined Kolmogorov
(1962) hypotheses). Non-Gaussianity of the conditional acceleration PDF is accounted for in terms of model nonlinearity.
The diffusion coefficient for the new model is chosen
based on DNS data for conditional two-time velocity statistics.
The resulting model predictions for conditional and unconditional velocity statistics and timescales
are shown to be in good agreement with DNS data.

\end{abstract}

%%%%%%%%%%%%%%%%%%%%%%%%%%%%%%%%%%%%%%%%%%%%%%%%%%%%%%%%%%%%%%%%%%%%%%%%%%%%%%%%%%%%%%%%%%%%%%%
%%%%%%%%%%%%%%%%%%%%%%%%%%%%%%%%%%%%%%%%%%%%%%%%%%%%%%%%%%%%%%%%%%%%%%%%%%%%%%%%%%%%%%%%%%%%%%%
%%%
%%%	SECTION: INTRODUCTION
%%%
%%%%%%%%%%%%%%%%%%%%%%%%%%%%%%%%%%%%%%%%%%%%%%%%%%%%%%%%%%%%%%%%%%%%%%%%%%%%%%%%%%%%%%%%%%%%%%%
%%%%%%%%%%%%%%%%%%%%%%%%%%%%%%%%%%%%%%%%%%%%%%%%%%%%%%%%%%%%%%%%%%%%%%%%%%%%%%%%%%%%%%%%%%%%%%%

\section{Introduction} \label{intro}
Of late, the statistics of fluid particle acceleration in turbulence have been the subject of many 
experimental (e.g., \cite{Vo98,Lap01,Vo02,Ch02,Gy04,Mo04}) and numerical (e.g., \cite{Ye97,Ve99,Bi05,Ye05}) efforts. 
These investigations have spurred a renewed interest (\cite{Po02,Be01,Be02,Re031}) in the modelling of conditional and unconditional statistics
of velocity and acceleration in terms of second-order Lagrangian stochastic models.
Recent work in this area has focused on the construction of stochastic models that are capable of reproducing
intermittency and Reynolds number effects in Lagrangian statistics as observed in experiments and DNS. In other words, 
second-order stochastic models can be formulated in such a way as to incorporate
accurate one-time statistics (and their Reynolds-number dependence) from experiments or DNS
and be able to reproduce intermittent two-time statistics in good agreement with experiments or DNS.

Reynolds-number effects in Lagrangian stochastic models were first addressed by \cite{S91}.
The Sawford (1991) model is exactly consistent with a joint-normal stationary one-time distribution for $\mathbf{Z} = [U,\,A]^T$,
where $U(t)$ and $A(t)$ denote (modelled) stochastic processes for one component of the Lagrangian velocity and acceleration.
As a result, the SDEs for the Sawford (1991) model are \emph{linear}:
\be
 d\mathbf{Z} = \begin{bmatrix} 0 & 1 \\ -{\sigma_A^2\over\sigma_U^2} &  -{b^2\over 2\sigma_A^2} \end{bmatrix}  \mathbf{Z} dt + \begin{bmatrix} 0 \\ b \end{bmatrix} d{W}    ,
\ee
where $\sigma_U$ and $\sigma_A$ denote standard deviations for velocity and acceleration, $b$ is a diffusion coefficient,
and $W$ is a standard Brownian motion (or Wiener process).
Sawford (1991) showed that matching of the second-order Lagrangian velocity structure function 
$D_U(s) = \langle(U(t+s) -U(t))^2 \rangle $ with the Kolmogorov (1941) hypotheses for the universal equilibrium range
uniquely identifies the diffusion coefficient as
\be
b= \sqrt{2\sigma_U^2 ({T_L^{\infty}}^{-1}+t_{\eta}^{-1}   ) {T_L^{\infty}}^{-1} t_{\eta}^{-1} }   ,
\ee
where $T_L^\infty = {2\over \mathcal{C}_0} {\sigma_U^2 \over \langle  \varepsilon \rangle }$ 
and $t_\eta = {\mathcal{C}_0 \over 2a_0 } \sqrt{{\nu \over \langle  \varepsilon \rangle }}$.
Here, $\mathcal{C}_0$ is the Kolmogorov constant for the second-order Lagrangian velocity structure function, 
$a_0$ is the acceleration variance normalized by the Kolmogorov scales, $\langle \varepsilon \rangle $ is the mean dissipation and
$\nu$ is the kinematic viscosity. 
Sawford (1991) also showed that model predictions are very close to DNS data for unconditional velocity and
acceleration autocorrelations at low Reynolds number. However, the Sawford model ignores intermittency 
of Lagrangian statistics and incorporates a Gaussian Lagrangian acceleration PDF, at variance with the observed
non-Gaussianity of acceleration found in experiments (\cite{Lap01}) and DNS (\cite{YP89,Ye05}).
%which is far from realityeven at low Reynolds number\cite{Lee05}.

\cite{Re031} addressed the problem of incorporating a strongly non-Gaussian PDF of acceleration into
a Lagrangian stochastic model.
Reynolds showed that an improved representation for the Lagrangian acceleration PDF in a second-order stochastic model
can be obtained by explicitly accounting for intermittency of dissipation. 
Specifically, he assumed a log-normal distribution for the dissipation rate, $\varepsilon$, together with a Gaussian assumption for the conditional PDF
of $A|\varepsilon$.
The latter assumption may be restated in terms of the \emph{conditionally standardized acceleration} defined by
$\tilde{A} \equiv {A\over \sigma_{A|\varepsilon}} $ (which has
zero and unit values for its conditional mean and variance).
In the Reynolds model, the conditional distribution $\tilde{A}|\varepsilon$ is assumed to be \emph{universal}
and, in particular, standard normal. This may be interpreted to imply that intermittency of 
dissipation is solely responsible for intermittency in acceleration.

Reynolds also assumed Gaussian velocity statistics and independence of velocity from dissipation and acceleration. In other words,
the Reynolds model is (by construction) exactly consistent with a joint-normal stationary one-time distribution of $(U,\tilde{A},\ln \varepsilon)$.

To completely specify his model, Reynolds assumed the Kolmogorov (1962) prediction for the conditional acceleration variance,
\be
\sigma_{A|\varepsilon}^2 / a_\eta^2 = a_0^* (\varepsilon / \langle \varepsilon\rangle)^{3/2}  ,
\ee 
where $a_0^*$ is a Kolmogorov constant and $a_\eta  = ( \langle \varepsilon\rangle^3 / \nu )^{1/4}$ is the Kolmogorov acceleration scale.
Following \cite{PC90}, Reynolds also assumed an Ornstein-Uhlenbeck (OU) process for $\chi \equiv \ln\varepsilon$.
The resulting model can be written as an SDE for $\mathbf{Z}=[ U,  \tilde{A} ,\ln \varepsilon - \langle\ln \varepsilon \rangle    ]^T$:
\be
d\mathbf{Z} = \begin{bmatrix} 0 & \sigma_{A|\varepsilon} & 0  \\ -{\sigma_{A|\varepsilon}\over\sigma_U^2} &  
 -{b^2\over 2\sigma_{A|\varepsilon}^2} & 0 \\ 0 & 0 & -T_\chi^{-1} \end{bmatrix}  \mathbf{Z} dt + 
 \begin{bmatrix} 0 & 0 \\ {\scriptstyle b/\sigma_{A|\varepsilon}} &0 \\ 0 & {\scriptscriptstyle \sqrt{2\sigma_\chi^2/T_\chi}  } \end{bmatrix} 
  \begin{bmatrix} dW \\ dW'  \end{bmatrix}      .
\label{re03model}
\ee
In these equations, $\sigma_\chi$ and $T_\chi$ denote the standard deviation and the integral scale for $\chi$, whereas $W$ and $W'$ are
independent Wiener processes.
The dissipation equation is effectively decoupled from the rest of the system and therefore the
Reynolds model is \emph{linear} in $U$ and $\tilde{A}$.
Additional assumptions made by Reynolds are: (i) a choice of diffusion coefficient made by analogy with 
the Sawford (1991) model, i.e.,
\be
b= \sqrt{2\sigma_U^2 ({T^{-1}_{L,\varepsilon}}+t_{\eta,\varepsilon}^{-1}   ) {T^{-1}_{L,\varepsilon}} t_{\eta,\varepsilon}^{-1} }   ,
\label{s91difcof}
\ee
where $T_{L,\varepsilon}= {2\over C_0} {\sigma_U^2 \over   \varepsilon  }$ and $t_{\eta,\varepsilon} = {C_0 \over 2a_0^* } \sqrt{{\nu \over   \varepsilon  }}$ 
($C_0$ being a model constant), and (ii) $T_\chi  \langle\varepsilon \rangle /\sigma_U^2=const$.

In this paper, we first review the basis for the Reynolds model against DNS. Then, a novel stochastic model is constructed
that incorporates one-time information from DNS and yields model predictions for two-time velocity statistics in good
agreeement with DNS.

The plan for this paper is as follows. %%In this paper we first review the basis for the Reynolds model against DNS.
In Section \ref{dnsdata} we review DNS data (first presented in \cite{Ye05}) for intermittency of dissipation, 
the PDF of conditionally standardized acceleration 
and the variance of acceleration conditioned on the pseudo-dissipation.
In Section \ref{ccgmodel}, a novel stochastic model that incorporates non-Gaussian one-time statistics from DNS is formulated.
Non-Gaussianity of the conditionally standardized acceleration PDF is accounted for in terms of nonlinearity in the model.
In Section \ref{specofb}, we show a choice of diffusion coefficient based on DNS data for conditional velocity autocorrelations that yields
model predictions for conditional and unconditional velocity autocorrelations and timescales in good agreement with DNS.
Conclusions for this work are summarized in Section \ref{concls}.

\section{DNS Data for Stochastic Modelling}
\label{dnsdata}

\subsection{Intermittency of Dissipation}

\begin{figure}
\centering %\vspace{2cm}
\subfigure{\label{cpdf2048}\epsfig{figure=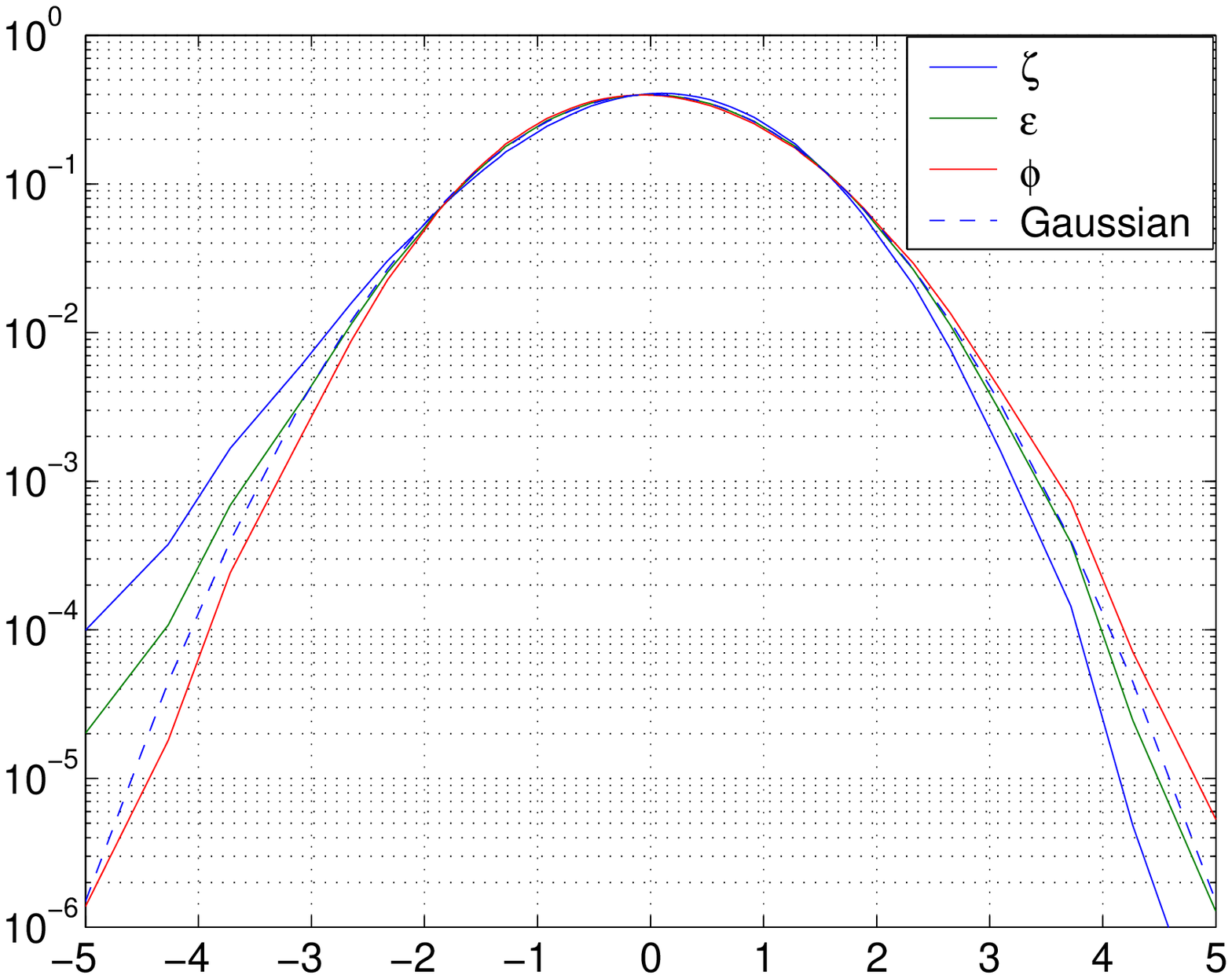,width=10.35cm}}
\caption{One-time PDFs (solid)
for the standardized logarithm of $X$ (with $X=\varepsilon, \, \zeta ,\, \varphi$) vs.~standard normal PDF (dashed).}
%%% add on to the figures; this is a trial and error process
\unitlength=1mm
\begin{picture}(0,0)(0,0)
    \put(-10,103){\scalebox{1}{  $ R_\lambda \approx 680$   }}
    \put(-62,55){\rotatebox{90}{\scalebox{1}{  PDF      }}}
    \put(-15,20){ \scalebox{1}{$ (\ln X -\langle \ln X \rangle) / \sigma_{\ln X}  $  } }
%   \put(-80,16){\vector(1,1){15}}
%   \put(-63,31){\scalebox{0.6}{$\gamma=0.2,\,0.185,\,0.18$}}
%   \put(-40,70){\scalebox{0.8}{$ C = 1.5 $}}
%   \put(-83,8){\scalebox{0.6}{$\gamma=0.3$}}
\end{picture}
\end{figure}

Figure \ref{cpdf2048} shows (one-time) PDFs for $\ln \varepsilon $, $\ln \zeta$ and $\ln \varphi$ for $R_\lambda \approx 680$
($R_\lambda =\sqrt{ {15\sigma_U^4 \over \nu \langle \varepsilon\rangle }  } $ being
the Taylor-scale Reynolds number), 
where $\varepsilon = 2\nu s_{ij}s_{ij}$ is the dissipation rate, $\zeta = 2\nu r_{ij}r_{ij}$ is the ``enstrophy'',
and $\varphi = \nu u_{i,j} u_{i,j}$ is the pseudo-dissipation ($s_{ij} $ and $r_{ij}$ being the strain-rate and rotation-rate
tensors, i.e., $u_{i,j} = s_{ij} + r_{ij}$). This figure suggests that pseudo-dissipation (as opposed to the dissipation rate, or the enstrophy)
is closest to log-normal for $R_\lambda \approx 680$. In fact, DNS data support this conclusion for Reynolds numbers
in the range $R_\lambda \approx 140-680$. 
In this paper we do not purport to discuss the validity of the log-normal model as opposed to more accurate intermittency
models for dissipation. We limit ourselves to the observation that pseudo-dissipation may be approximately described
as log-normal for $R_\lambda \approx 140-680$.

\begin{figure}
\centering %\vspace{2cm}
\subfigure{\label{cf2eps1024}\epsfig{figure=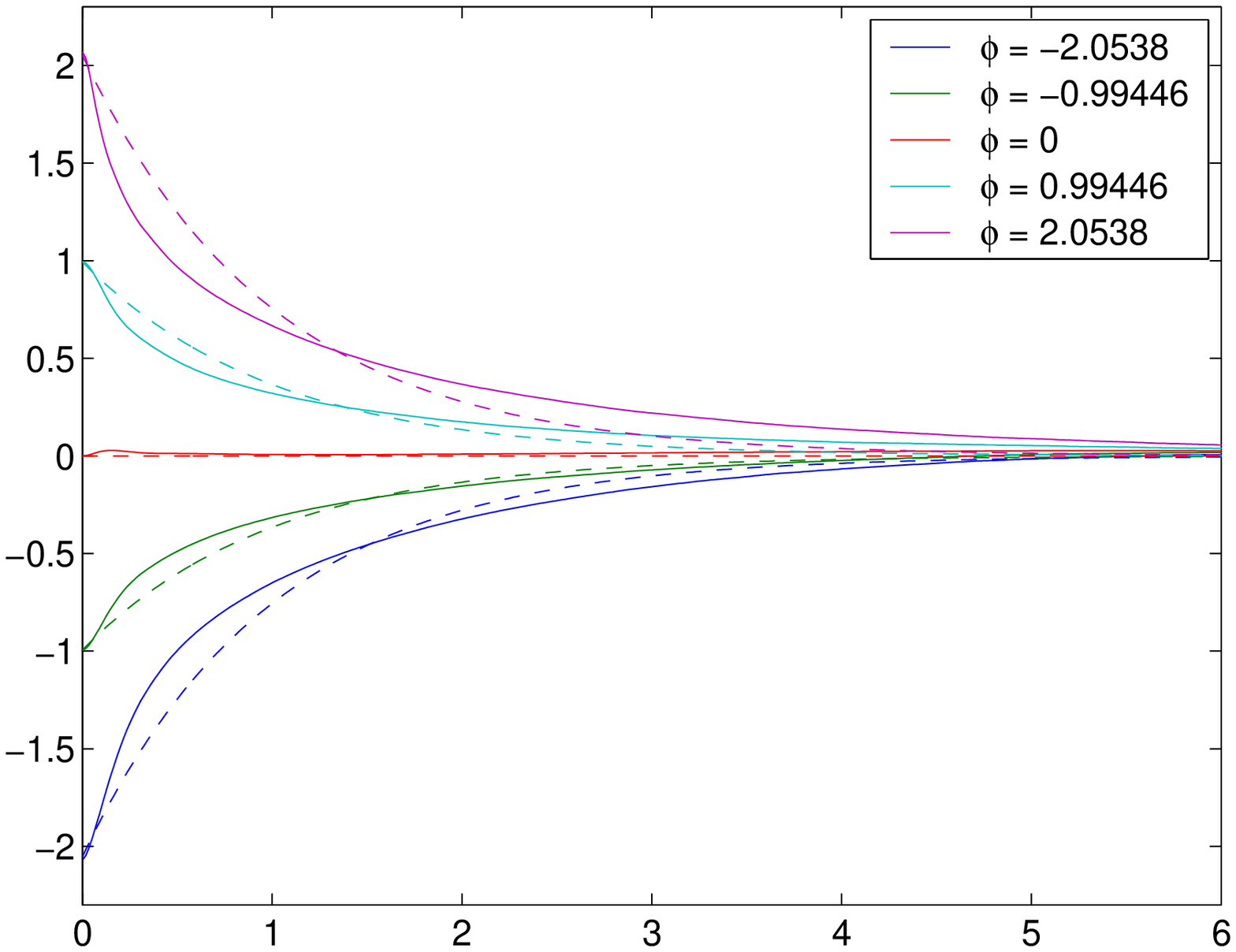,width=10.35cm}}
\caption{DNS data for conditional expectations (solid) with 
$X=(\ln \varepsilon - \langle \ln \varepsilon\rangle)/\sigma_{\ln \varepsilon}  $ vs.~$\phi \,e^{-\tau/T_X}$ (dashed,
$T_X$ being the Lagrangian integral timescale for $X$).}
%%% add on to the figures; this is an trial and error process
\unitlength=1mm
\begin{picture}(0,0)(0,0)
    \put(-10,103){\scalebox{1}{  $ R_\lambda \approx 400    $   }}
    \put(-59,47){\rotatebox{90}{\scalebox{1}{  $ \langle X(t+\tau) | X(t)=\phi \rangle  $     }}}
    \put(-8,21){ \scalebox{1}{ $  \tau / T_X $} }
%   \put(-90,71){\scalebox{0.6}{ $\circ\!- \;\;\;N_{sim}=16$ }}
%    \put(-56,-4){$\kappa_d$}
%    \put(-80,105){\scalebox{0.6}{$t=0$}}
\end{picture}
%%% if you do not want to write anything to the figure, you can remove this part
\\
\subfigure{\label{cf2psds1024}\epsfig{figure=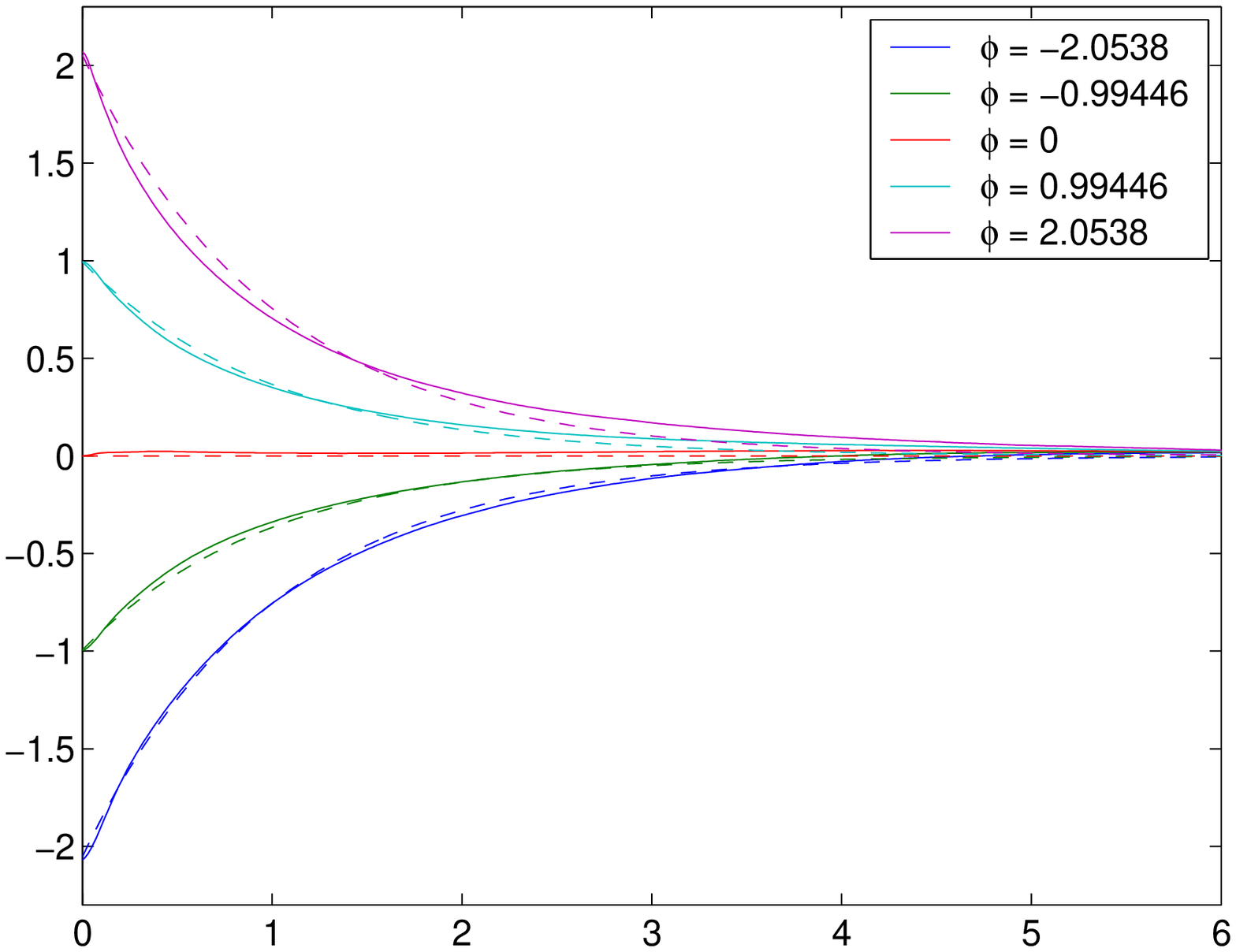,width=10.35cm}}
\caption{DNS data for conditional expectations (solid) with $X=(\ln \varphi - \langle \ln \varphi\rangle)/\sigma_{\ln \varphi} $
vs.~$\phi \,e^{-\tau/T_X}$~(dashed, $T_X$ being the Lagrangian integral timescale for $X$).}
%%% add on to the figures; this is an trial and error process
\unitlength=1mm
\begin{picture}(0,0)(0,0)
    \put(-10,103){\scalebox{1}{  $ R_\lambda \approx 400    $   }}
    \put(-59,47){\rotatebox{90}{\scalebox{1}{  $ \langle X(t+\tau) | X(t)=\phi \rangle  $     }}}
    \put(-8,20){ \scalebox{1}{ $  \tau / T_X $} }
%   \put(-90,71){\scalebox{0.6}{ $\circ\!- \;\;\;N_{sim}=16$ }}
%    \put(-56,-4){$\kappa_d$}
%    \put(-80,105){\scalebox{0.6}{$t=0$}}
\end{picture}
%%% if you do not want to write anything to the figure, you can remove this part
\end{figure}

Let us now assume that 
$\chi \equiv \ln \varphi/\langle \varepsilon\rangle $ evolves by an OU process,
\be
d\chi = -\left(  \chi +{\sigma_\chi^2 \over 2}\right) {dt\over T_\chi} + \sqrt{{2\sigma_\chi^2 \over T_\chi} } dW'  .
\label{chipsds}
\ee
It follows that $\langle X(t) | X(0)=\phi \rangle  =\phi e^{-t /T_\chi}$,  
where $X\equiv (\ln \varphi -\langle \ln\varphi\rangle)/\sigma_{\ln \varphi} $.
Figure \ref{cf2psds1024}  % \ref{cf2psds256}--\ref{cf2psds1024} 
shows (Lagrangian) two-time conditional means of $X$ from DNS 
(Fig.~\ref{cf2eps1024} shows two-time conditional means from DNS with $X$ re-defined in terms of $\varepsilon $).
These figures suggest that two-time conditional means of the (standardized) logarithm of pseudo-dissipation are closest to
simple exponentials (at least for $R_\lambda \approx 140-680$, based on DNS at different Reynolds numbers).
Therefore, on the basis of DNS, we argue that, for the purposes of stochastic modelling, 
pseudo-dissipation is well-approximated by an OU process.

To completely specify the dissipation model \eqref{chipsds} (where $\chi \equiv \ln \varphi/ \langle \varepsilon\rangle $), $\sigma_\chi $ and $T_\chi$ have to be prescribed.
As discussed by \cite{Ye05}, the DNS data support the Kolmogorov (1962) prediction for
\be
\sigma_{\chi}^2 = A + {3\mu \over 2} \ln R_\lambda   ,
\ee
with $\mu \approx 0.25$, in good agreement with \cite{SK93} ($A\approx -0.863$
is reported by \cite{Ye05}).
The integral timescale $T_\chi $ is chosen to match the DNS data
(Table \ref{Tchitab}).

\begin{table}
$$
\begin{array}{|c||c|c|c|c|c|}\hline
 R_\lambda & 38 & 140 & 240 & 400 & 680  \\ \hline\hline 
T_\chi / T & 0.3317  &  0.1719  &  0.1311 &   0.1056  &  0.0906 \\ \hline
\end{array}
$$
\caption{Values of $T_\chi / T  $ (with $T\equiv 1.5 \sigma_U^2 / \langle\varepsilon \rangle$) from DNS of homogeneous turbulence.}
\label{Tchitab}
\end{table}

\subsection{PDF of Conditionally Standardized Acceleration}

\begin{figure} \label{figure14}
\centering \vspace{-1cm}
\epsfig{bbllx=150, bblly=250, bburx=590,bbury=570, clip=,
figure=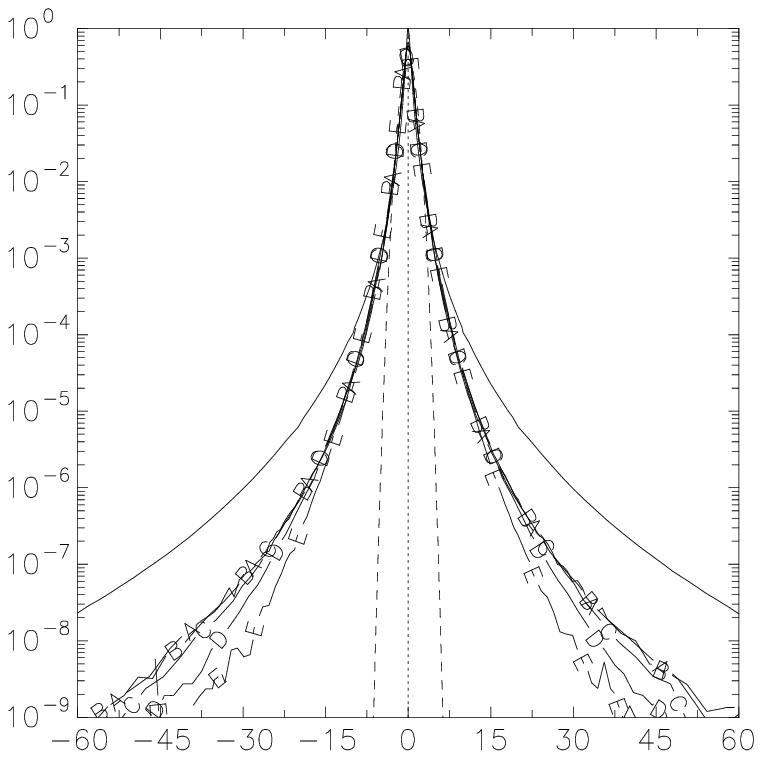,angle=-90,width=15cm} 
\vspace{-8cm}
\caption{Standardized onditional PDFs of $\tilde{A}$ given $\varepsilon= \hat{\varepsilon}$ from $2048^3$ DNS at $R_\lambda \approx 680$. Lines A-E are for
$\hat{\varepsilon} / \langle \varepsilon  \rangle = \{ 0.0359, \,0.136,\, 0.469,\, 1.62,\, 6.05\}$, corresponding to 
$Z=(\ln \hat{\varepsilon} - \langle \varepsilon\rangle  )/\sigma_{\ln\varepsilon } = \{ -2.054, \, -0.994, \, 0,\, 0.994,\, 2.054\} $. Also shown are
the unconditonal PDF of acceleration (solid unmarked line) and a standard Gaussian PDF (dashed).  }      
\unitlength=1mm
\begin{picture}(0,0)(0,0)

\put(-60,88){\rotatebox{90}{ \scalebox{1}{ PDF  }}}
\put(-8,37){ \scalebox{1}{ $ a/ \sigma_{A|\hat{\varepsilon}}, \;\; a/\sigma_{A} $}} 
\end{picture}
\end{figure}

\begin{figure}  \label{figure15}
\centering \vspace{-1cm}
\epsfig{bbllx=150, bblly=250, bburx=590,bbury=570, clip=,
figure=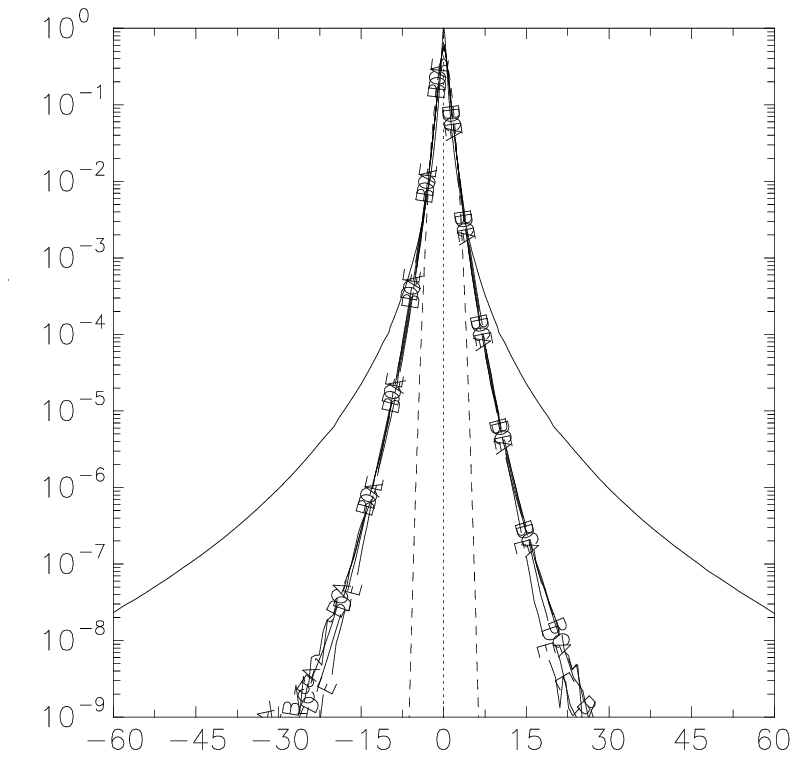,angle=-90,width=15cm}
\vspace{-8cm}
\caption{Conditional PDFs of $\tilde{A}$ given $\varphi= \hat{\varphi}$ from $2048^3$ DNS at $R_\lambda \approx 680$. Lines A-E are for
$\hat{\varphi} / \langle \varphi  \rangle = \{ 0.0362, \,0.134,\, 0.458,\, 1.56,\, 5.79\}$, corresponding to 
$Z=(\ln \hat{\varphi} - \langle \varphi\rangle  )/\sigma_{\ln\varphi } = \{ -2.054, \, -0.994, \, 0,\, 0.994,\, 2.054\} $. Also shown are
the unconditonal PDF of acceleration (solid unmarked line) and a standard Gaussian PDF (dashed).  } 
\unitlength=1mm
\begin{picture}(0,0)(0,0)
\put(-60,88){\rotatebox{90}{ \scalebox{1}{ PDF  }}}
\put(-8,37){ \scalebox{1}{ $ a/ \sigma_{A|\hat{\varphi}}, \;\; a/\sigma_{A} $}} 

\end{picture}
\end{figure}

We now investigate the conditionally Gaussian assumption in the Reynolds model.
Figure 4 shows PDFs $f_{\tilde{A}|\varepsilon}(a|\hat{\varepsilon})$ for different values of $\hat{\varepsilon}$
for $R_\lambda \approx 680$ (a standard normal PDF and the unconditional acceleration PDF 
for $R_\lambda \approx 680$ are also shown).
Figure 5 shows analogous information when $\varphi$ is used in place of $\varepsilon$.
Both sets of conditional PDFs are much less intermittent (with weaker tails at large
fluctuations) than the unconditional acceleration PDF. This is particularly true of the PDFs of $\tilde{A}|\varphi$
(with $\tilde{A} \equiv A/\sigma_{A|\varphi}$) which, however, still show significant non-Gaussian behaviour.
A remarkable degree of collapse of these PDFs for different values of the conditioning variable is notable (except for
very small and very large conditional fluctuations). Therefore, to a first approximation, the PDFs
$f_{\tilde{A}|\varphi}(\tilde{a}|\hat{\varphi})$ may be described as approximately independent of $\hat{\varphi}$. In fact,
based on simulations at different Reynolds numbers (not shown), the conditional PDF of $\tilde{A}|\varphi$ may also be 
(approximately) described as independent of the Reynolds number.
 
\begin{figure}
\centering %\vspace{2cm}
\subfigure{\label{clinlog}\epsfig{figure=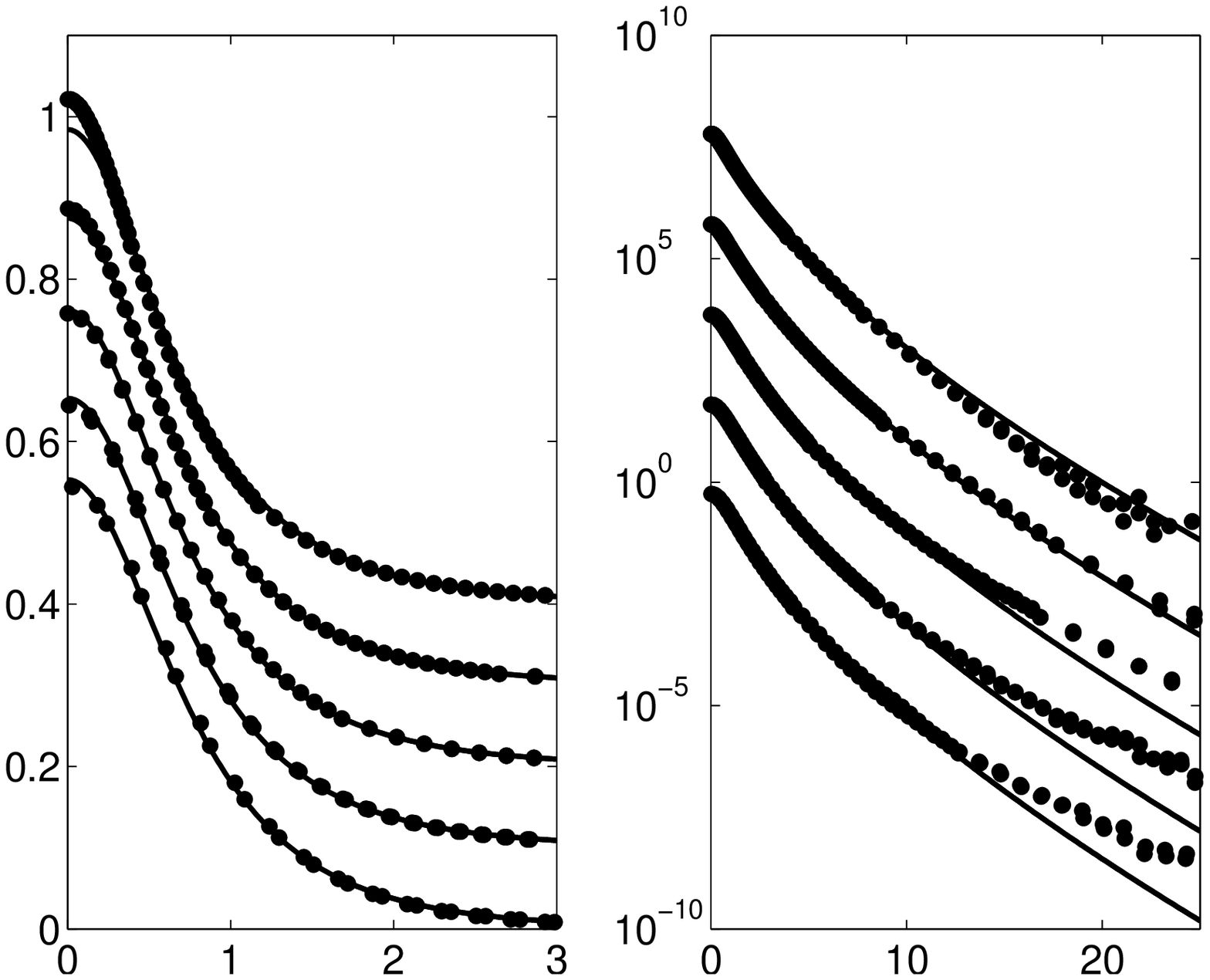,width=10.35cm}}
\caption{Standardized conditional PDFs of $\tilde{A}$ given $\varphi= \hat{\varphi}$. The symbols are from the $2048^3$ DNS; the lines are
the cubic-Gaussian PDF with the same kurtosis as the DNS data. The values of the conditioning variable are such that
$Z=(\ln \hat{\varphi} - \langle \varphi\rangle  )/\sigma_{\ln\varphi } = \{ -2.054, \, -0.994, \, 0,\, 0.994,\, 2.054\} $. 
In each plot, the lower curve and the $y$-axis correspond to the lowest conditioning value. The curves for the other
conditioning values are successively shifted upwards, by an amount $0.2$ in the linear plot (left), and by a factor of $100$ in the
logarithmic plot (right).  } %%% add on to the figures; this is an trial and error process
\unitlength=1mm
\begin{picture}(0,0)(0,0)
 %%   \put(-10,98){\scalebox{1}{  $ R_\lambda \approx 400    $   }}
    \put(-59,92){\rotatebox{90}{\scalebox{1}{  PDF      }}}
    \put(-30,56){ \scalebox{1}{ $ |a| / \sigma_{A|\varphi}$} }
	\put(20, 56){ \scalebox{1}{ $ |a| / \sigma_{A|\varphi}$} }
%   \put(-90,71){\scalebox{0.6}{ $\circ\!- \;\;\;N_{sim}=16$ }}
%    \put(-56,-4){$\kappa_d$}
%    \put(-80,105){\scalebox{0.6}{$t=0$}}
\end{picture}
\end{figure}

\cite{Ye05} suggest that the PDF of $\tilde{A}|\varphi$ can be described (to a very good approximation) as cubic-Gaussian.
By definition, a random variable $Z$ is cubic-Gaussian with parameter $p$ (also denoted as $Z\sim G^3(p)$) if
\be
Z = C[(1-p) X + pX^3] ,
\ee
where $X$ is a standardized Gaussian random variable and $C$ is determined by the standardization condition
$\langle Z^2 \rangle =1$ as $C(p) = (1+4p+10p^2)^{-1/2}$. Figure \ref{clinlog} shows that the cubic-Gaussian PDF
provides a remarkably accurate description of the conditional PDFs $f_{\tilde{A}|\varphi}(\tilde{a}|\hat{\varphi})$.
Comparable accuracy to that in Fig.~\ref{clinlog} is achieved when the cubic-Gaussian PDF is used to fit
DNS data for $f_{\tilde{A}|\varphi}(\tilde{a}|\hat{\varphi})$ at lower Reynolds numbers 
(not shown, but see Sect.~IV of \cite{Ye05}).
A value of $p\approx 0.1$ results from the observation (based on DNS) that $\mu_4(\tilde{A}|\hat{\varphi}) \approx 8$
(approximately independent of $\hat{\varphi}$ and $R_\lambda$; more details in \cite{Ye05}).

\begin{figure}
\centering %\vspace{2cm}
\subfigure{\label{condavar}\epsfig{figure=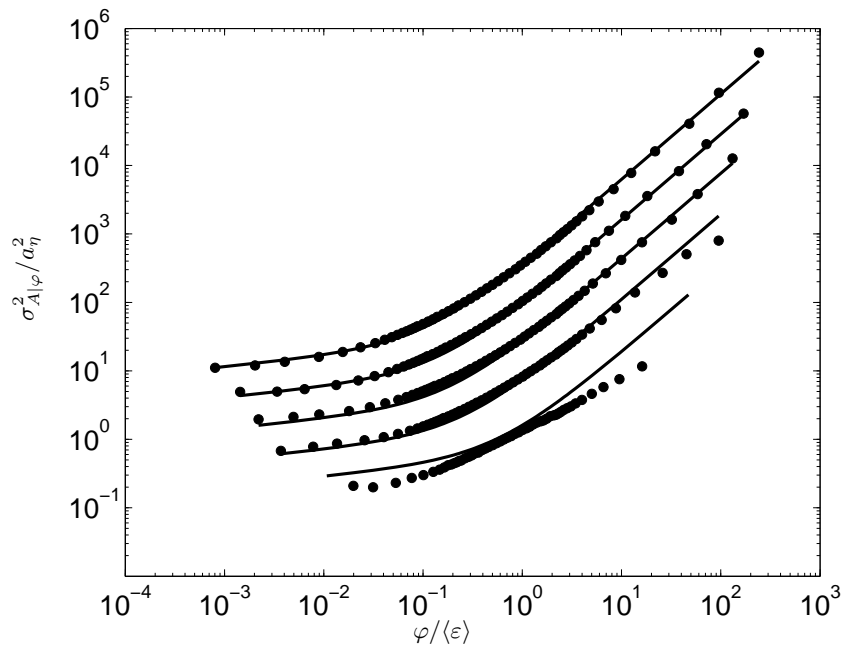,width=10.35cm}}
\caption{Variance of acceleration conditioned on the pseudo-dissipation
for different values of $R_\lambda$. The symbols are the DNS data; the lines are the empirical fit Eq.~\eqref{popesfit}.
The lowest curve and the $y$-axis correspond to $R_\lambda \approx 38$. The  other four curves are for 
$R_\lambda \approx 140, \, 240, \, 400, \, 680$, successively shifted upwards by a factor of $\sqrt{10}$. } %%% add on to the figures; this is an trial and error process
\unitlength=1mm
\begin{picture}(0,0)(0,0)
 %%   \put(-10,98){\scalebox{1}{  $ R_\lambda \approx 400    $   }}
    \put(-59,75){\rotatebox{90}{\scalebox{1}{   $ \sigma_{A|\varphi}^2 / a_\eta^2  $     }}}
    \put(-8,34){ \scalebox{1}{ $  \varphi / \langle \varepsilon \rangle  $} }
%%	\put(20, 39){ \scalebox{1}{ $ |a| / \sigma_{A|\varphi}$} }
%   \put(-90,71){\scalebox{0.6}{ $\circ\!- \;\;\;N_{sim}=16$ }}
%    \put(-56,-4){$\kappa_d$}
%    \put(-80,105){\scalebox{0.6}{$t=0$}}
\end{picture}
\end{figure}

\subsection{Conditional Acceleration Variance}
With a view to the joint (stochastic) modelling of acceleration and pseudo-dissipation, 
we now investigate the validity of the Kolmogorov (1962) prediction for the conditional acceleration variance,
\be
\sigma_{A|\varphi}^2 / a_\eta^2 = a_0^* (\varphi/ \langle \varepsilon \rangle )^{3/2}   .
\label{sigaphi62}
\ee
Figure \ref{condavar} shows $\varphi$-dependences from DNS for the conditional acceleration variance at different Reynolds numbers.
In the same figure, the following expression (first presented in \cite{Ye05})
\be
{ \sigma_{A|\varphi}^2  \over a_\eta^2}   = 
{1.2\over R_\lambda^{0.2}} \left( {\varphi\over \langle \varepsilon\rangle }  \right)^{0.15} 
 + \ln\left({R_\lambda\over 20} \right) \left( {\varphi\over \langle \varepsilon\rangle }  \right)^{1.25}  ,
 \label{popesfit}
\ee
is shown to be an accurate representation (except at the smallest $R_\lambda$) of the DNS data.
As may be seen, the low-$\varphi$
behaviour for the conditional acceleration variance deviates strongly from that predicted by Eq.~\eqref{sigaphi62}.
Also, careful measurement of the slope
for the large-$\varphi$ portion of the curves in Fig.~\ref{condavar} yields values that are systematically
less than $1.5$, again at variance with the Kolmogorov (1962) prediction.

Equation \eqref{popesfit} is most useful for stochastic modelling purposes because it accurately parameterizes the
conditional acceleration variance (given $\varphi$) in terms of both the value being conditioned upon
and the Reynolds number.

\section{Conditionally Cubic-Gaussian (CCG) Stochastic Lagrangian Models}
\label{ccgmodel}
Lagrangian statistics for $\varepsilon$, $\zeta$ and $\varphi$ from DNS show that stochastic modelling 
is easiest when pseudo-dissipation is used in place of the dissipation rate or the enstrophy.
This is because (i) $\varphi$ is closest to log-normal, and (ii)  two-time conditional means of $\ln \varphi$
are closest to exponential, and (iii) the conditional PDFs of acceleration given $\varphi =\hat{\varphi} $ collapse
best, and with the narrowest tails. Thus, we base the model on pseudo-dissipation $\varphi$
and take the OU process Eq.~\eqref{chipsds} as its stochastic model.

Conditioning on pseudo-dissipation is most useful when considering the joint-statistics 
of acceleration and pseudo-dissipation because the PDF of $\tilde{A}|\varphi$
may be described (to a first approximation) as \emph{universal} and, in particular, cubic-Gaussian.
In other words, given $\varphi=\hat{\varphi}$ and a standardized Gaussian random variable $\bar{A}$,
the acceleration $A$ can be modelled as 
\be
A=\sigma_{A|\hat{\varphi}} C[(1-p)\bar{A} + p \bar{A}^3]  .
\label{atiltox}
\ee
For given $\varphi$, the relation between $A$ and $\bar{A}$ is invertible (and one-to-one).

The stochastic model is most conveniently expressed in terms of the velocity $U(t)$, the ``Gaussian'' acceleration $\bar{A}(t)$
(related to the acceleration by Eq.~\eqref{atiltox}), and the log-pseudo-dissipation variable
$\chi^* \equiv \chi -\langle \chi\rangle = \ln (\varphi / \langle \varepsilon\rangle  )- \langle \ln (\varphi/  \langle \varepsilon\rangle)  \rangle$.

The model is 
\begin{gather}
dU = Adt = \sigma_{A|\varphi}C [ (1-p) \bar{A} + p\bar{A}^3] dt ,  \label{ueqn} \\
d\bar{A}  = \bar{\theta}  dt + \bar{b} dW  ,      \label{abareqn} \\
d\chi^*  = -\chi^* {dt\over T_\chi} +  \sqrt{{2\sigma_\chi^2 \over T_\chi}}  dW'  ,  \label{chieqn}
\end{gather}
where $\bar{\theta}$ and $\bar{b}$ are drift and diffusion coefficients specified below.

The stationary one-time joint PDF of $U,\,\bar{A}$ and $\chi^*$ is denoted by $f(v,\bar{a},x^*)$,
where $v,\bar{a}$ and $x^*$ are sample-space variables corresponding to $U,\bar{A}$ and $\chi^*$.
We now assume that this PDF is joint-normal with the variables being uncorrelated with each other
(at the same time). Thus, with the assumptions made the joint PDF is 
\be
f= {1\over \sigma_U \sqrt{2\pi }} \exp\left( -{ v^2 \over 2\sigma_U^2 }  \right) 
{1\over  \sqrt{2\pi }} \exp\left( -{ \bar{a}^2 \over 2 } \right)
 {1\over \sigma_\chi \sqrt{2\pi }} \exp\left( -{ {x^*}^2 \over 2\sigma_\chi^2} \right)    .
\label{asspdf}
\ee
The imposition of this PDF leads to a constraint for the drift coefficient $\bar{\theta} $ in Eq.~\eqref{abareqn}, namely,
\be
\bar{\theta}(v,\bar{a},\varphi)  =  -{\sigma_{A|\varphi}\over \sigma_U^2 } C v (1+p+p \bar{a}^2) + 
{\bar{b}^2 \over 2} {\partial \over \partial \bar{a}} \ln \bar{b}^2 f + \bar{\theta}^*    ,  
\ee
where $\bar{\theta}^*$ is any function such that ${\partial\over \partial \bar{a}}( \bar{\theta}^*f)=0$, which for simplicity
we take to be zero.
We also introduce the assumption $\partial \bar{b}/\partial \bar{a}=0$ (which can be supported using an adiabatic elimination 
argument in the limit $R_\lambda \to \infty$). Then, Eq.~\eqref{abareqn} can be rewritten as
\begin{equation}
d\bar{A}  = -{\bar{b}^2 \over 2 }\bar{A} dt-{\sigma_{A|\varphi}\over \sigma_U^2}U C (1+p  + p\bar{A}^2)   dt 
+ \bar{b} dW   . \\
\end{equation}
This equation (together with Eqs.~\eqref{ueqn} and \eqref{chieqn}) defines a class of CCG models, i.e., different models
with the same stationary distribution \eqref{asspdf}
correspond to different choices of $\bar{b}$.
Each model captures the conditional dependence of acceleration on pseudo-dissipation based on DNS
(Eq.~\eqref{popesfit}) that accounts for deviations from the Kolmogorov (1962) hypotheses. Also, each model is non-linear because
it accounts for the non-Gaussianity of the conditionally standardized
acceleration PDF.

\section{Specification of Diffusion Coefficient}
\label{specofb}

\begin{figure}
\centering %\vspace{2cm}
\subfigure{\label{u03vdns700}\epsfig{figure=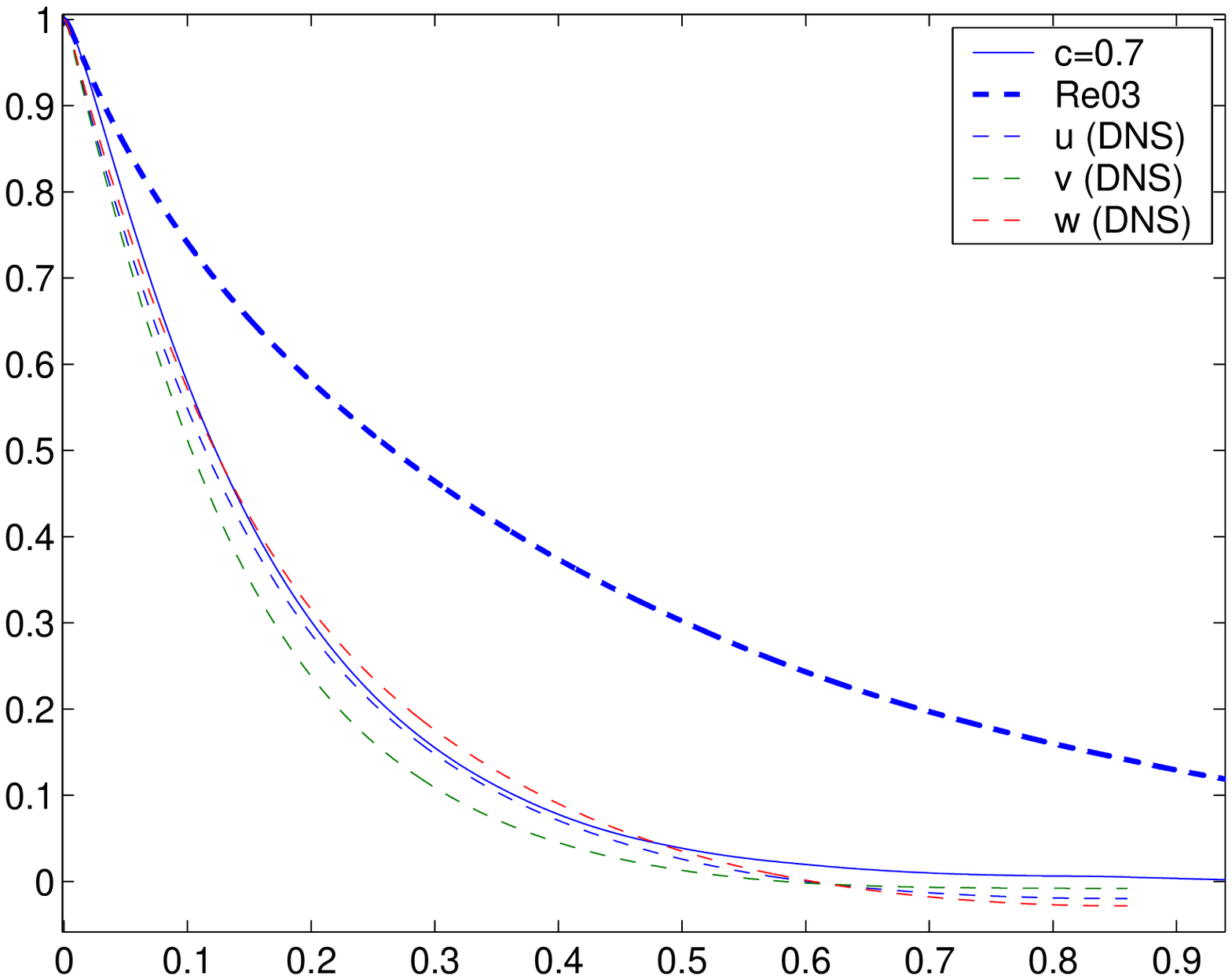,width=10.35cm}}
\caption{Velocity autocorrelation for CCG model (solid) vs.~DNS data (dashed). The thicker, dashed line is for the Reynolds (2003) model.}
%%% add on to the figures; this is a trial and error process
\unitlength=1mm
\begin{picture}(0,0)(0,0)
    \put(-10,106){\scalebox{1}{  $ R_\lambda \approx 680   $   }}
    \put(-59,40){\rotatebox{90}{\scalebox{1}{   Velocity Autocorrelation   }}}
    \put(-8,18){ \scalebox{1}{$  t /T   $  } }
%   \put(-80,16){\vector(1,1){15}}
%   \put(-63,31){\scalebox{0.6}{$\gamma=0.2,\,0.185,\,0.18$}}
%   \put(-40,70){\scalebox{0.8}{$ C = 1.5 $}}
%   \put(-83,8){\scalebox{0.6}{$\gamma=0.3$}}
\end{picture}
\end{figure}

\begin{figure}
\centering %\vspace{2cm}
\subfigure{\label{Tb2oxmp1}\epsfig{figure=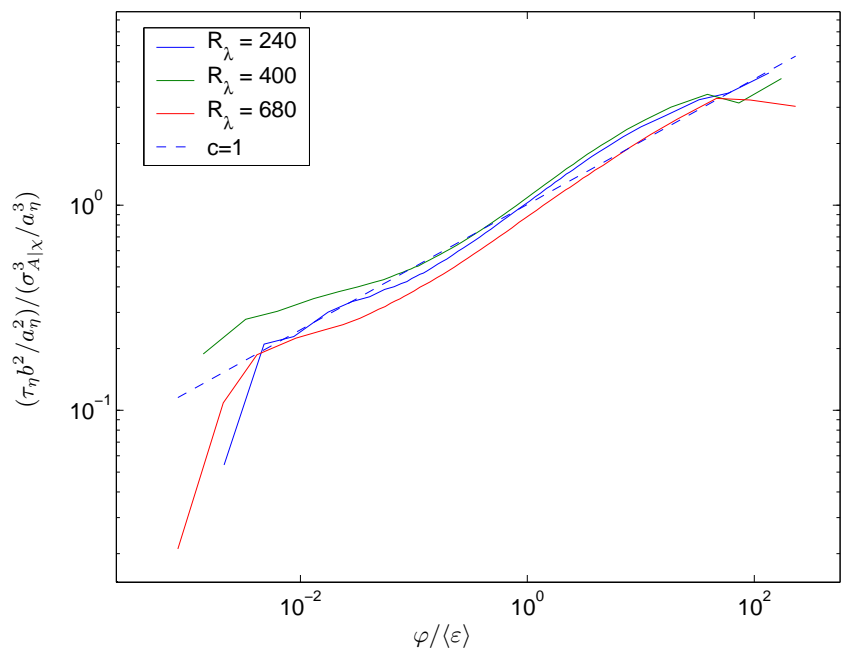,width=10.35cm}}
\caption{Phi-dependences for diffusion coefficient (solid) for the Reynolds (2003) model with a frozen dissipation that matches the DNS
values of $\sigma_{A|\chi} $ and $\tau_{U|\chi}$. The dashed line is a tentative fit to the solid curves
$G(\varphi) = \left( {\varphi \over \langle \varepsilon \rangle }  \right)^{0.3}$ (Eq.~\eqref{difcof0}).} %%% add on to the figures; this is an trial and error process
\unitlength=1mm
\begin{picture}(0,0)(0,0)
 %%   \put(-10,98){\scalebox{1}{  $ R_\lambda \approx 400    $   }}
    \put(-59,57){\rotatebox{90}{\scalebox{1}{    $ (\tau_\eta b^2 / a_\eta^2) / (\sigma_{A|\chi}^3 / a_\eta^3) $    }}}
    \put(-8,27){ \scalebox{1}{ $  \varphi / \langle \varepsilon \rangle  $} }
%%	\put(20, 39){ \scalebox{1}{ $ |a| / \sigma_{A|\varphi}$} }
%   \put(-90,71){\scalebox{0.6}{ $\circ\!- \;\;\;N_{sim}=16$ }}
%    \put(-56,-4){$\kappa_d$}
%    \put(-80,105){\scalebox{0.6}{$t=0$}}
\end{picture}
\end{figure}

\begin{figure}
\centering %\vspace{2cm}
\subfigure{\label{uuc720p4}\epsfig{figure=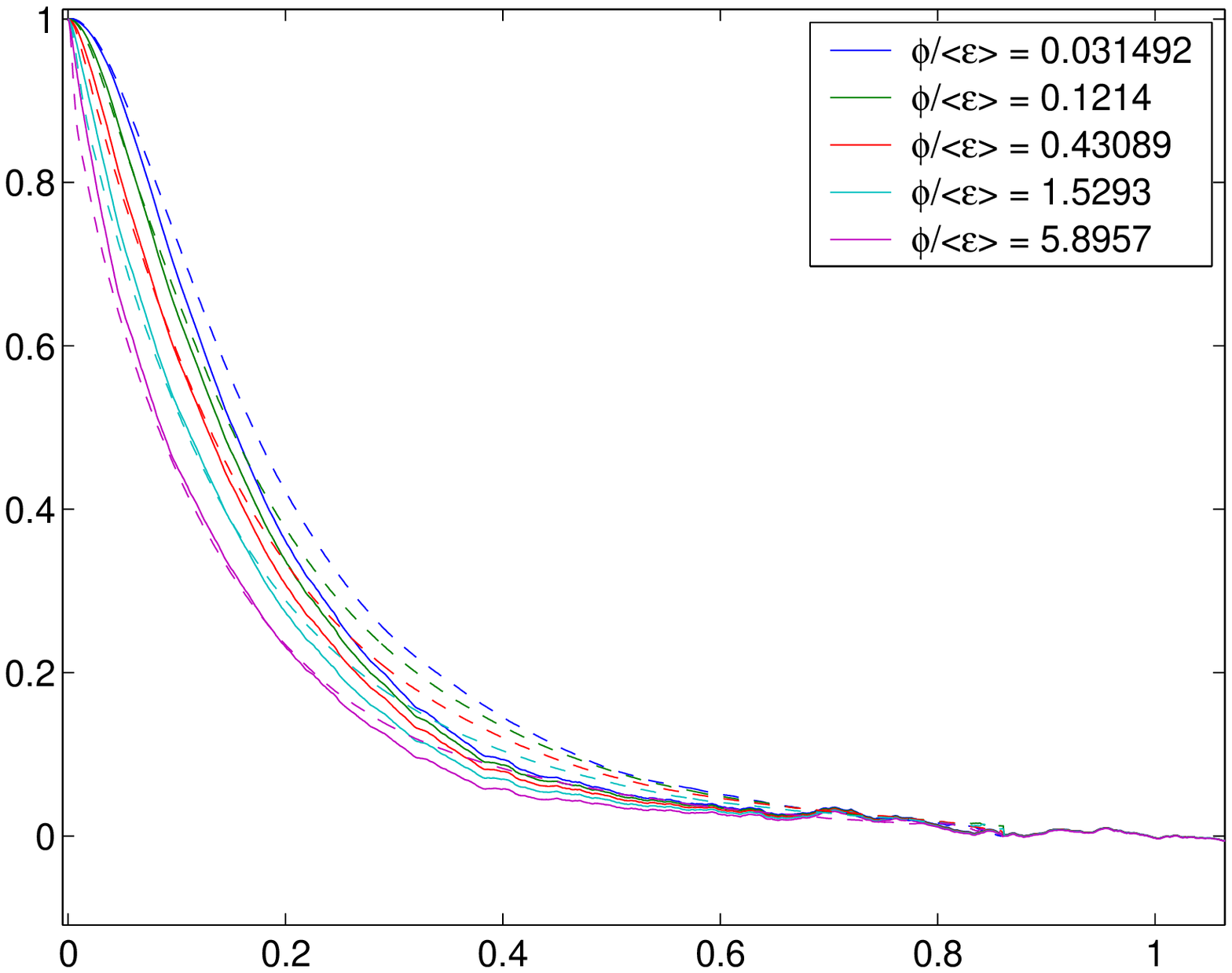,width=10.35cm}}
\caption{Conditional velocity autocorrelations for CCG model (solid) vs.~DNS (dashed).}
%%% add on to the figures; this is an trial and error process
\unitlength=1mm
\begin{picture}(0,0)(0,0)
    \put(-10,98){\scalebox{1}{  $ R_\lambda \approx 680    $   }}
    \put(-59,40){\rotatebox{90}{\scalebox{1}{ Velocity autocorrelations     }}}
    \put(-8,12){ \scalebox{1}{ $  t / T  $} }
%   \put(-90,71){\scalebox{0.6}{ $\circ\!- \;\;\;N_{sim}=16$ }}
%    \put(-56,-4){$\kappa_d$}
%    \put(-80,105){\scalebox{0.6}{$t=0$}}
\end{picture}
\end{figure}

\begin{figure}
\centering %\vspace{2cm}
\subfigure{\label{tauuchi720mp1}\epsfig{figure=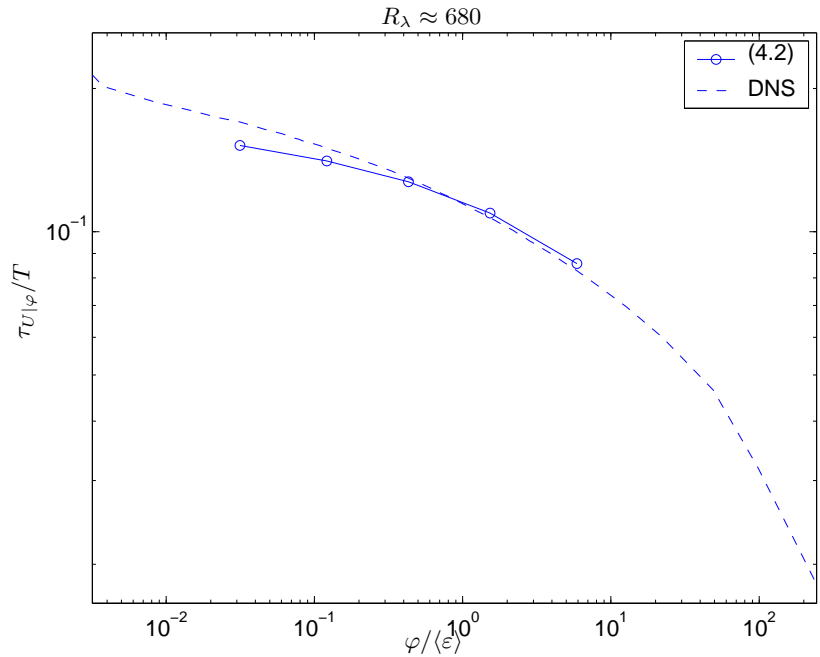,width=10.35cm}}
\caption{Model predictions for halving times for conditional velocity autocorrelations (solid) vs.~DNS (dashed).}
\unitlength=1mm
\begin{picture}(0,0)(0,0)
 %%   \put(-10,98){\scalebox{1}{  $ R_\lambda \approx 400    $   }}
    \put(-8,105){\scalebox{1}{  $ R_\lambda \approx 680   $   }}
    \put(-56,62){\rotatebox{90}{\scalebox{1}{   $ \tau_{U|\varphi} / T $  }}}
    \put(-6,22){ \scalebox{1}{ $  \varphi / \langle\varepsilon  \rangle  $  } }
%%%	\put(20, 39){ \scalebox{1}{ $ |a| / \sigma_{A|\varphi}$} }
%   \put(-90,71){\scalebox{0.6}{ $\circ\!- \;\;\;N_{sim}=16$ }}
%    \put(-56,-4){$\kappa_d$}
%    \put(-80,105){\scalebox{0.6}{$t=0$}}
\end{picture}
\end{figure}

\begin{figure}
\centering %\vspace{2cm}
\subfigure{\label{aa720p4x07}\epsfig{figure=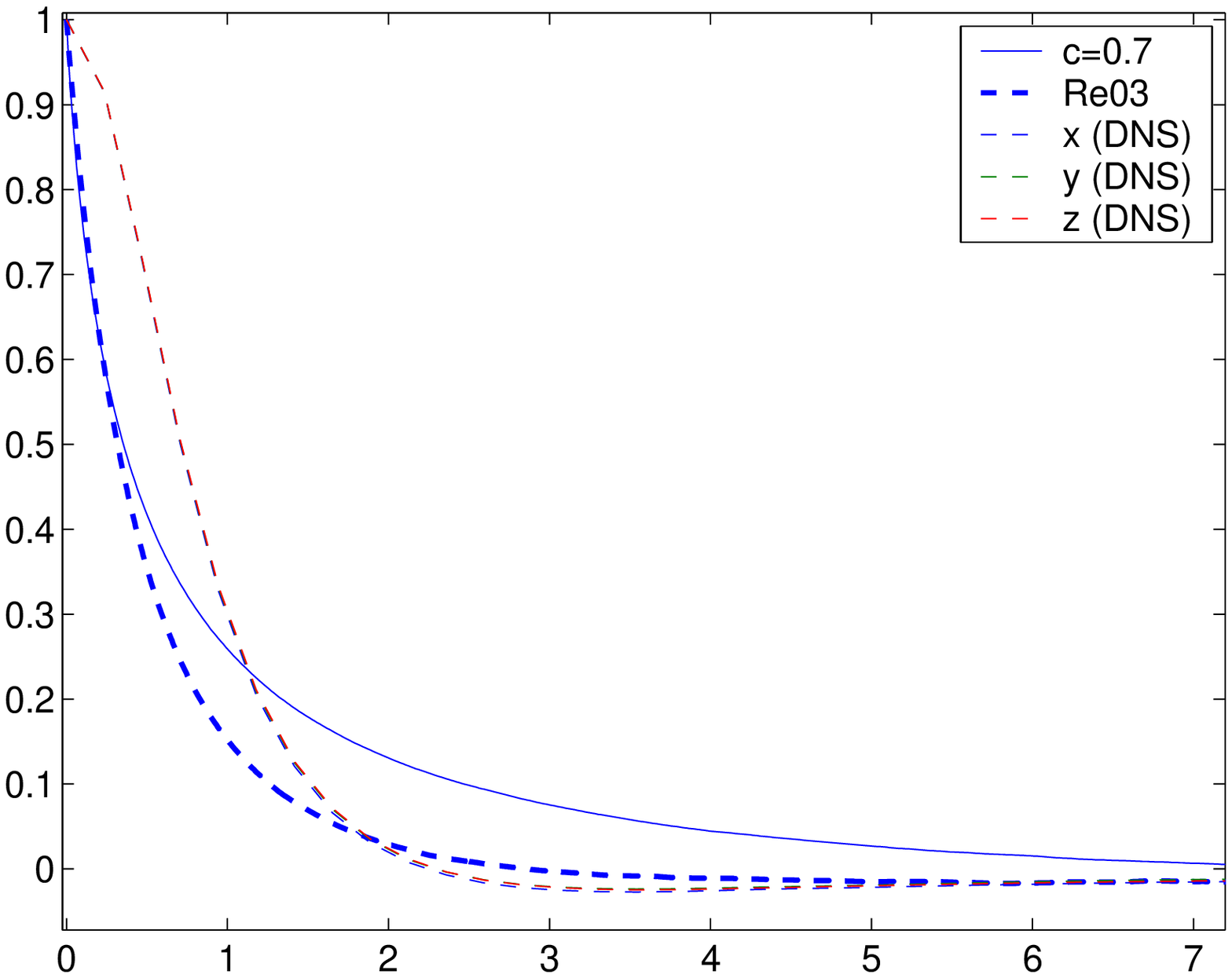,width=10.35cm}}
\caption{Acceleration autocorrelation for CCG model (solid) vs.~DNS (dashed). The thicker, dashed line is for the Reynolds (2003) model.}
\unitlength=1mm
\begin{picture}(0,0)(0,0)
 %%   \put(-10,98){\scalebox{1}{  $ R_\lambda \approx 400    $   }}
    \put(-10,106){\scalebox{1}{  $ R_\lambda \approx  680    $   }}
    \put(-59,40){\rotatebox{90}{\scalebox{1}{   Acceleration Autocorrelation   }}}
    \put(-8,20){ \scalebox{1}{$  t / \tau_\eta   $  } }
%%	\put(20, 39){ \scalebox{1}{ $ |a| / \sigma_{A|\varphi}$} }
%   \put(-90,71){\scalebox{0.6}{ $\circ\!- \;\;\;N_{sim}=16$ }}
%    \put(-56,-4){$\kappa_d$}
%    \put(-80,105){\scalebox{0.6}{$t=0$}}
\end{picture}
%%% if you do not want to write anything to the figure, you can remove this part
\\
\subfigure{\label{aac720p4}\epsfig{figure=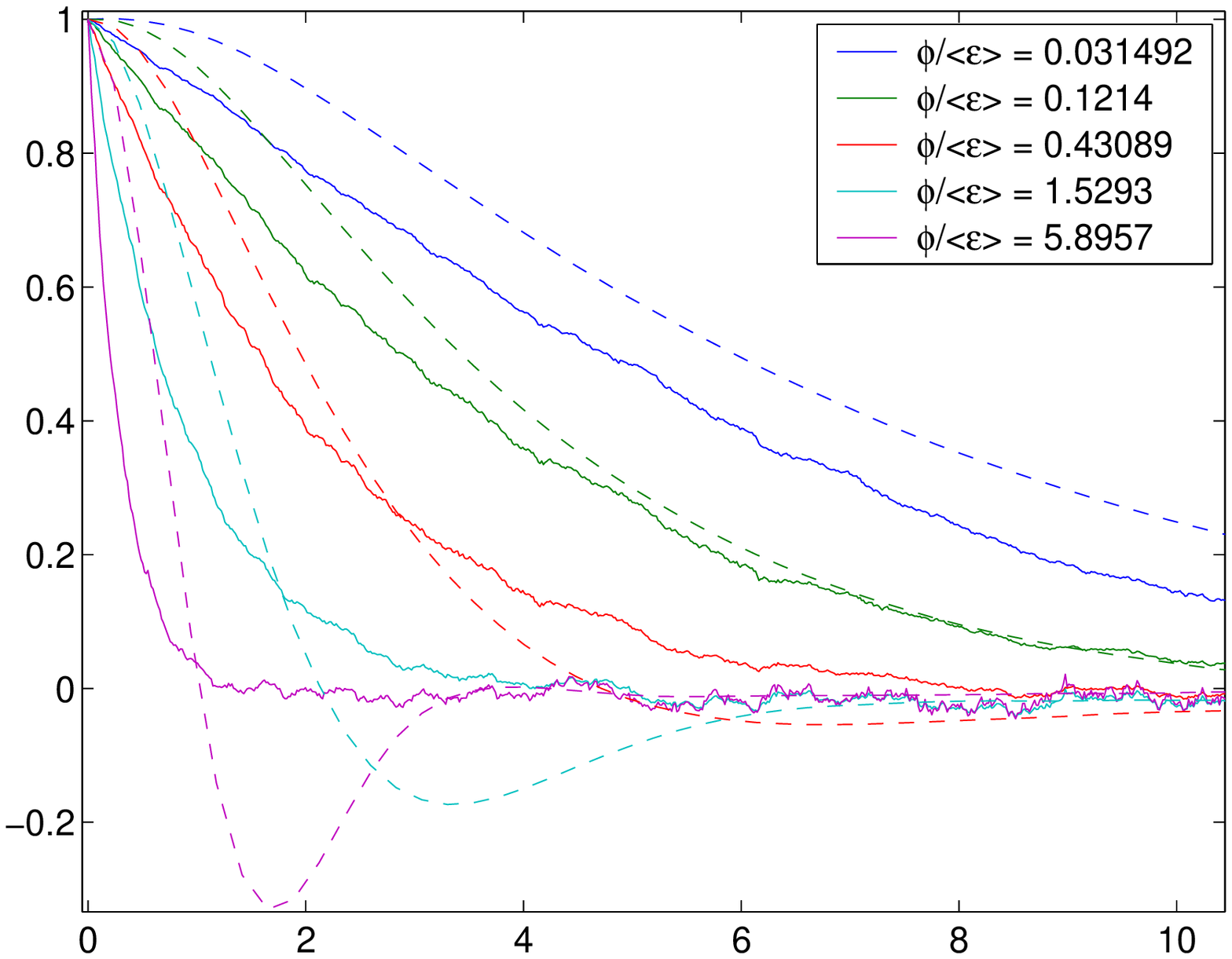,width=10.35cm}}
\caption{Conditional acceleration autocorrelations for CCG model (solid) vs.~DNS (dashed).}
%%% add on to the figures; this is a trial and error process
\unitlength=1mm
\begin{picture}(0,0)(0,0)
    \put(-10,98){\scalebox{1}{  $ R_\lambda \approx 680   $   }}
    \put(-59,40){\rotatebox{90}{\scalebox{1}{   Acceleration autocorrelations   }}}
    \put(-8,12){ \scalebox{1}{$  t / \tau_\eta  $  } }
%   \put(-80,16){\vector(1,1){15}}
%   \put(-63,31){\scalebox{0.6}{$\gamma=0.2,\,0.185,\,0.18$}}
%   \put(-40,70){\scalebox{0.8}{$ C = 1.5 $}}
%   \put(-83,8){\scalebox{0.6}{$\gamma=0.3$}}
\end{picture}
%%% if you do not want to write anything to the figure, you can remove this part
\end{figure}

\begin{figure}
\centering %\vspace{2cm}
\subfigure{\label{dupdf720p4}\epsfig{figure=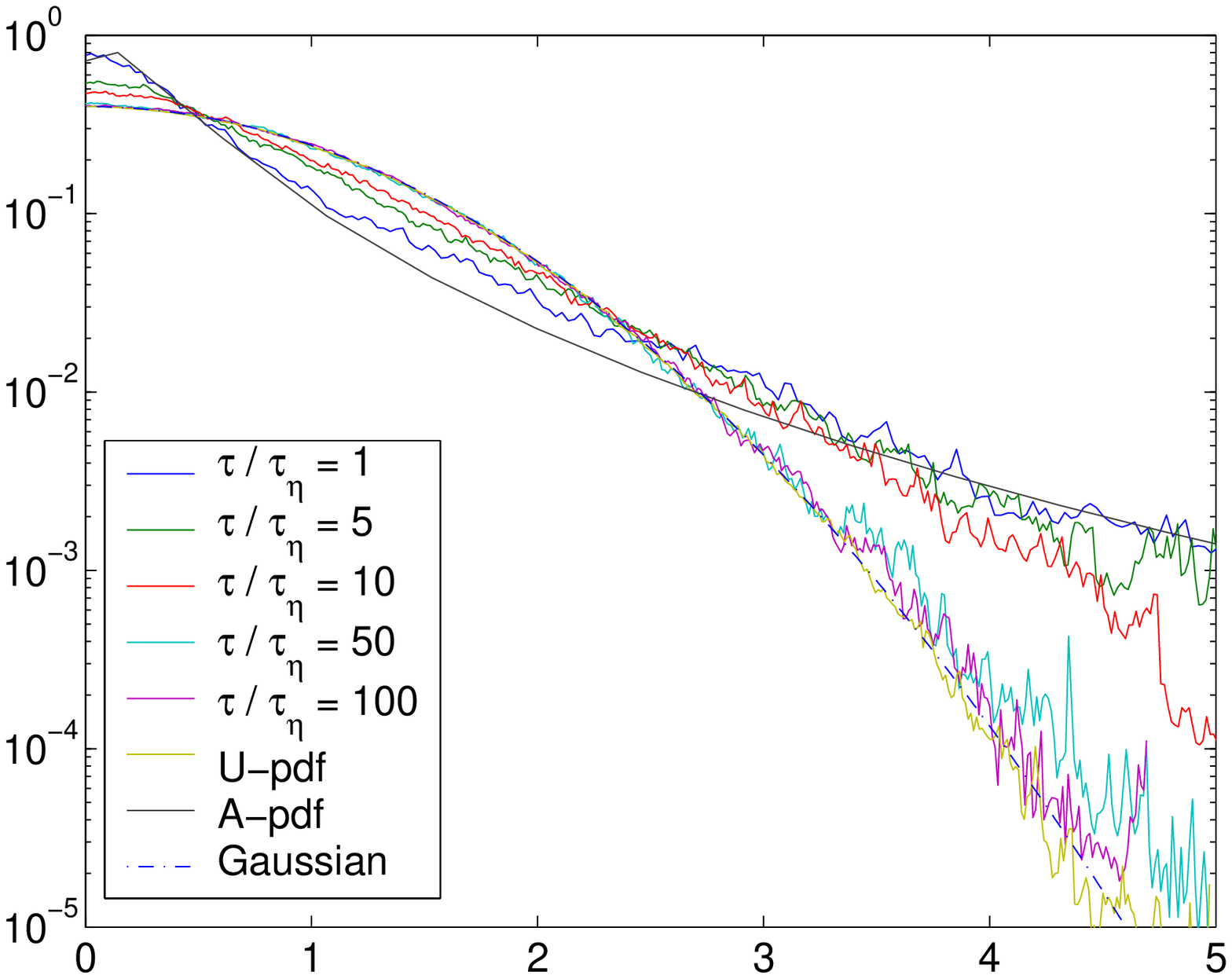,width=10.35cm}}
\caption{Model predictions for standardized PDFs of Lagrangian velocity increments.}
\unitlength=1mm
\begin{picture}(0,0)(0,0)
 %%   \put(-10,98){\scalebox{1}{  $ R_\lambda \approx 400    $   }}
    \put(-8,96){\scalebox{1}{  $ R_\lambda \approx  680    $   }}
    \put(-58,55){\rotatebox{90}{\scalebox{1}{   PDF   }}}
    \put(-11,14){ \scalebox{1}{$  v/\sigma_U , \; a/\sigma_A , \; \text{etc.}  $  } }
%%	\put(20, 39){ \scalebox{1}{ $ |a| / \sigma_{A|\varphi}$} }
%   \put(-90,71){\scalebox{0.6}{ $\circ\!- \;\;\;N_{sim}=16$ }}
%    \put(-56,-4){$\kappa_d$}
%    \put(-80,105){\scalebox{0.6}{$t=0$}}
\end{picture}
\end{figure}

The diffusion coefficient for the Reynolds (2003) model is specified by analogy with the Sawford (1991) model
(Eq.~\eqref{s91difcof}). This choice is arbitrary and, furthermore, it leads to poor agreement with DNS
for unconditional autocorrelations (Fig.~\ref{u03vdns700}).

We now investigate the question of how to select $\bar{b}(\chi)$
by reference to two-time conditional velocity statistics from DNS.
To this end, the Reynolds model with a frozen dissipation $\chi \equiv \hat{\chi}$ is considered. Then, the first two equations
in \eqref{re03model} revert to a Sawford '91 type of model. 
With the frozen model, we investigate $\chi$-dependences for $b$ such that the frozen model matches the DNS values of
$\sigma_{A|\chi}$ and $\tau_{U|\chi}$ ($\tau_{U|\chi }$ being halving times for
conditional velocity autocorrelations $\rho_{U|\chi}(t) \equiv \langle U(t) U(0) | \chi(0)=\chi \rangle  $,
i.e., $\rho_{U|\chi}(\tau_{U|\chi})=1/2$).
Results from such calculations for different values of $R_\lambda$ are shown in Fig.~\ref{Tb2oxmp1}.
This figure would suggest that $\varphi$-dependences for $(\tau_\eta b^2 / a_\eta^2 )/ ( \sigma_{A|\chi } ^3 / a_\eta^3)  $
at different Reynolds numbers may be approximately described in terms of a single function
$G(\varphi) = \left({\varphi \over \langle\varepsilon \rangle} \right)^{0.3} $.
The diffusion coefficient is then given by
\be
\left({b \over \sigma_{A|\chi}} \right)^2 =  {G(\varphi) } {\sigma_{A|\chi} \over u_\eta}  ,
\label{difcof0}
\ee
where $u_\eta =(\nu \langle\varepsilon\rangle )^{1/4}$ is the Kolmogorov velocity scale.
We then use this determination to specify $\bar{b}$ in \eqref{abareqn}, i.e.~we let
\be
\bar{b}^2 = c(R_\lambda) G(\varphi)  {\sigma_{A|\chi}  \over u_\eta}   ,
\label{difcof}
\ee
where $c(R_\lambda)$ is a correction factor (given in Table \ref{cofrlam}) so selected as to ensure 
good agreement between model predictions and DNS data
for unconditional velocity autocorrelations (Fig.~\ref{u03vdns700}). 
The resulting model predictions for conditional  
velocity autocorrelations and timescales are shown in Figs.~\ref{uuc720p4} and \ref{tauuchi720mp1}.   %%%Figs.~\ref{tauuchi233mp1}--\ref{tauuchi720mp1} vs.~DNS. 
Model predictions for unconditional and conditional
acceleration autocorrelations are shown in Figs.~\ref{aa720p4x07} and \ref{aac720p4}.
These plots suggest that the specification \eqref{difcof} is capable of reproducing reasonable agreement with DNS data
for conditional and unconditional velocity autocorrelations and timescales.
Also, Eq.~\eqref{difcof} yields PDFs of Lagrangian velocity increments that are approximately Gaussian for large time-lags
(Fig.~\ref{dupdf720p4}). These PDFs develop stretched tails as the time-lag decreases and ultimately approach the
Lagrangian acceleration PDF for very small time-lags. This behaviour is consistent with recent observations of Lagrangian intermittency
in experiments and simulations (\cite{Mo04}).

\begin{table}
$$
\begin{array}{|c||c|c|c|c|c|}\hline
 R_\lambda & 38 & 140 & 240 & 400 & 680  \\ \hline\hline 
c(R_\lambda) & 1.4  &  1.0  &  1.0 &   1.0  &  0.7 \\ \hline
\end{array}
$$
\caption{Values for $c(R_\lambda) $ in Eq.~\eqref{difcof}.}
\label{cofrlam}
\end{table}

\section{Conclusions} \label{concls}

After a brief review of the basis for the Reynolds (2003) model against DNS,
we have shown the formulation of a novel stochastic based on simple data assimilation from DNS.
This model is exactly consistent with Gaussian velocity
and conditionally cubic-Gaussian acceleration statistics and incorporates a representation for
the logarithm of pseudo-dissipation as an Ornstein-Uhlenbeck process.
The new model captures the effects of intermittency
of dissipation on acceleration and the deviations from the Kolmogorov (1962) hypotheses (based on DNS)
in the conditional dependence of acceleration on pseudo-dissipation.
%%with deviations from the Kolmogorov (1962) hypotheses accounted for.  
Further, non-Gaussianity of the conditionally standardized acceleration PDF (as observed in DNS)
is captured in terms of model nonlinearity.
An empirical specification of diffusion coefficient based on DNS data for conditional velocity timescales 
yields reasonable agreement with DNS data for conditional and unconditional velocity autocorrelations and
timescales.

\vskip 1cm
\noindent

\begin{acknowledgments}
We gratefully acknowledge support from the National Science Foundation through Grants 
No.~CTS-0328329 and CTS-0328314, with
computational resources provided by the Pittsburgh Supercomputing Center and the San Diego Supercomputer Center,
which are both supported by NSF.

\end{acknowledgments}

\bibliography{lagbib}
\bibliographystyle{jfm}

\end{document}